# Thermal characterization of convective heat transfer in microwires based on modified steady state "hot wire" method

Xiaoman Wang[1,2,#], Rulei Guo[1,2,#], Qinping Jian[1,2,#], Guilong Peng[1,2], Yanan Yue[3],Nuo Yang*[1,2]

[1]State Key Laboratory of Coal Combustion, Huazhong University of Science and Technology (HUST), Wuhan 430074, People's Republic of China
[2]Nano Interface Center for Energy (NICE), School of Energy and Power Engineering, Huazhong University of Science and Technology (HUST), Wuhan 430074, People's Republic of China
[3]School of Power and Mechanical Engineering, Wuhan University, Wuhan 430072, People's Republic of China

#XW, RG, and QJ contribute equally on this work.

*Corresponding author E-mail: nuo@hust.edu.cn (NY)

# ABSTRACT

The convection plays a very important role in heat transfer when MEMS work under air environment. However, traditional measurements of convection heat transfer coefficient require the knowledge of thermal conductivity, which makes measurements complex. In this work, a modified steady state "hot wire" (MSSHW) method is proposed, which can measure the heat transfer coefficient of microwires' convection without the knowledge of thermal conductivity. To verify MSSHW method, the convection heat transfer coefficient of platinum microwires was measured in the atmosphere, whose value is in good agreement with values by both traditional measurement methods and empirical equations. Then, the convection heat transfer coefficient of microwires with different materials and diameters were measured by MSSHW. It is found that the convection heat transfer coefficient of microwire is not sensitive on materials, while it increases from 86 $W/(m^2 \cdot K)$ to 427 $W/(m^2 \cdot K)$ with the diameter of microwires decreasing from 120 μm to 20 μm. Without knowing thermal conductivity of microwires, the MSSHW method provides a more convenient way to measure the convective effect.

**INTRODUCTION**

With the prosperous development of micro-electro-mechanical systems (MEMS), heat dissipation within highly integrated circuits has drawn wide attention[1]. The chip-level heat generated by the increasing power density is pushing the demand of better cooling methods[2]. If the trend of integration and miniaturization keeps following the International Technology Roadmap for Semiconductors (ITRS), thermal management will become the bottleneck for further development of the electronic devices[2-4]. In the past two decades, the heat transfer in one-dimensional (1-D) structures has been investigated, which are widely employed in MEMS. It is important for understanding the thermal properties of materials[5,6], enhancing heat transfer in MEMS theoretically and experimentally [7-9], and enriching applications in energy conversion, such as nanowire solar cell[10,11] and thermoelectric application[12,13].

According to previous works, there are several methods to measure the thermal conduction of 1-D microwires including the steady-state hot wire method [14], the $3\omega$ method [15-17] and the Raman spectroscopy method [18-20]. Among these methods, the steady-state hot wire method is one of the most convenient and simplest method to measure the thermal conduction of 1-D structure. To exclude convective heat transfer effect, these approaches are normally carried out in a vacuum environment.

Besides the thermal conduction, convective heat transfer has also been investigated. It has been proved that the convection shows different performances when the size of sample shrinks to micro/nanoscale. For example, convection heat transfer coefficient continuously increases from $5\sim10\,\text{W}/(\text{m}^2 \cdot \text{K})$ (natural convection on bulk material) [21] to over $4000\,\text{W}/(\text{m}^2 \cdot \text{K})$ (micro carbon fiber with a diameter of 4.3 μm) [22].Moreover, convection heat transfer coefficient of carbon nanotubes with a diameter of 1.47 nm can reach $8.9 \times 10^4\,\text{W}/(\text{m}^2 \cdot \text{K})$ in an atmosphere environment [23]. Such high convection heat transfer coefficient under micro/nanoscale indicates that the convection plays a significant role in heat dissipation in MEMS devices.

There are three main methods to measure the convection heat transfer coefficient of 1-D structure. The first one is the modified 3ω method[24]. By adding convection part into the original model, the convection heat transfer coefficient can be measured when other thermal properties are known, including thermal conductivity. The second one is Raman mapping measurement. By comparing Raman temperature measurement result with the value from resistance change by joule heating, Zhang's group searched for the optical absorption coefficient and then use the best fitting result to get the convection heat transfer coefficient during iteration[22]. The third one is taking advantage of the steady-state method to analyze the air side[25,26]. However, this method ignores the heat conduction within the sample. Theoretical calculation has shown that the heat conduction can be overlooked only when the length of the tested sample has reached a certain value[23].

Since conductive and convective effects are coupled in heat transfer of MEMS and traditional methods ignore thermal conduction or get it by other ways, a method for characterizing convective heat transfer effect without the knowledge of thermal conduction is convenient and important. Yue et al. proposed a method to study conductive and convective heat transfer of microwires simultaneously by using steady-state Joule-heating and Raman mapping together[7]. However, the expensive Raman equipment and the relatively large signal error limit its broader application.

In this paper, we proposed a modified steady state hot wire (MSSHW) method which can characterize convection heat transfer coefficient of microwires without the knowledge of thermal conductivity. Firstly, we theoretically derived the mathematical model of MSSHW method. Secondly, we measured convection heat transfer coefficient of a platinum microwire to verify MSSHW method. Then, using MSSHW method, we studied the material dependence of convection heat transfer coefficient by measuring microwires of platinum, stainless steel and tungsten. Lastly, microwires with different diameters ranging from 20 to 120 μm were measured to study the size effect on convection heat transfer coefficient. This work presents a convenient experimental method, MSSHW method, to study the heat transfer at microscale by convection.

**MODEL & METHOD**

The schematic illustration of experimental setup is shown in Fig. 1a. The sample is connected to a multimeter (Keithley 2700) and a current Source (Keithley 6221) in a four-electrodes configuration. A direct current is applied on the two outside electrodes and the corresponding voltage is measured on the two inside electrodes, which diminishes the contact electric resistance. The photo of experimental setup is shown in Fig. 1b.

When a direct current is applied to an electroconductive microwire, the microwire will heat itself up due to Joule heating. The temperature increase is closely related to both the conduction and the convection of the microwire, when the radiation can be neglected. When a series of microwires with different length has been measured, the thermal conductivity can be eliminated from the calculation of convection heat transfer coefficient. Thus, it provides a possibility to measure the convection heat transfer coefficient without knowing thermal conductivity.

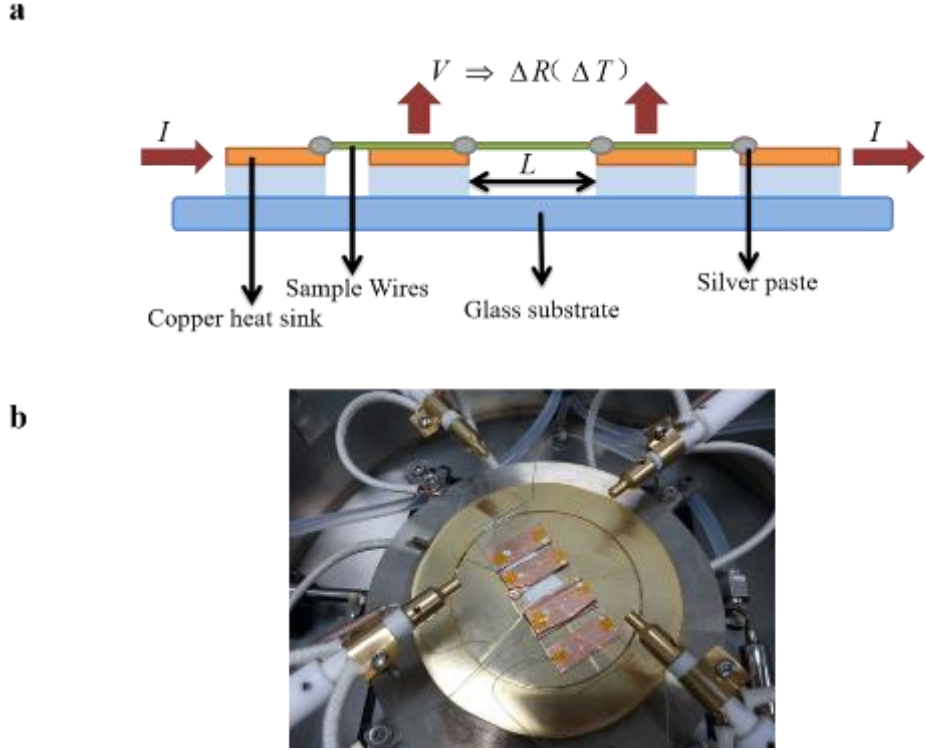

**Fig.1** (a) The sketch schematic illustration of experimental setup with a four-electrodes configuration (not to scale). (b) Photo of experimental setup.

The heat transfer in a microwire can be described by 1-D heat transfer equation as:

$$\kappa A \frac{d^2 T(x)}{dx^2} - hP(T(x) - T_E) - \varepsilon \sigma P(T^4(x) - T_E^4) + qA = 0 \qquad (1)$$

where $\kappa$ is the thermal conductivity, $A$ is the cross area, $h$ is the convection heat transfer coefficient, $T_E$ is the environment temperature, $P$ is the perimeter, $\varepsilon$ is the emissivity, $\sigma$ is the Stefan-Boltzmann constant as $5.67 \times 10^{-8} \, W/(m^2 \cdot K^4)$, and $q$ is the heat generated per unit volume. The four terms in the left side represent the thermal conduction, the natural convection, the radiation and the heat generated inside the thin wire, respectively.

The contribution from radiation is much smaller than that from conduction and convection, which can be therefore neglected. To confirm this assumption, the temperature profiles were calculated with/without radiation according to Eq. 1 by MATLAB. The calculated sample is a platinum microwire, whose length and diameter are 16.3 mm and 32.6 μm, respectively [24]. The size of the calculated sample is a typical situation in the real experiments. By applying $\kappa = 66.5\,W/(m \cdot K)$, $h = 410\,W/(m^2 \cdot K)$ [24] and $\varepsilon = 0.075$ into Eq. 1 [27], we can get the temperature profile of the wire as shown in Fig. 2a. We calculated the average temperature of four microwires with different lengths as shown in Fig. 2b, and radiation shows negligible influence on the temperature profile of those microwires.

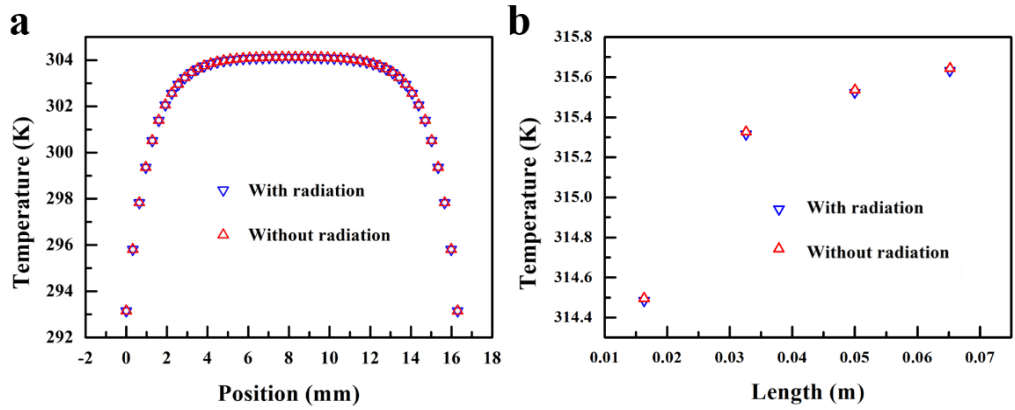

**Fig.2** (a) The temperature distribution for platinum wire with d=32.6 μm, l=16.3 mm under the current of 60 mA with and without radiation effect considered. (b). Average temperatures for platinum wires with different lengths (d=32.6 μm). Blue down triangles represent values when radiation effect is considered. Red up triangles represent the values when radiation effect is ignored

Thus, the 1-D heat transfer equation can be simplified as

$$\kappa A \frac{d^2 T(x)}{dx^2} - hP(T(x) - T_E) + qA = 0 \qquad (2)$$

Since the heat sinks have much larger heat capacities compared to the thin wire, the boundary condition is thus expressed as

$$\begin{cases} T(0) = T_E \\ T(L) = T_E \end{cases} \qquad (3)$$

where L is the length of the microwire.

The solution of the 1-D heat transfer equation is

$$T(x) = T_E + \frac{q}{km^2} + \frac{\left(-\frac{q}{km^2}e^{2mL} + \frac{q}{km^2}e^{mL}\right)e^{-mx} + \left(-\frac{q}{km^2}e^{mL} + \frac{q}{km^2}\right)e^{mx}}{e^{2mL} - 1} \qquad (4)$$

$$m = \sqrt{\frac{hP}{\kappa A}}$$

Then, $T(x)$ is integrated over the length and the average temperature increase is calculated as

$$\Delta T = \frac{1}{L} \int_0^L T(x) dx - T_E = \frac{q}{\kappa m^2} \left(1 - \frac{2(e^{mL}-1)}{mL(e^{mL}+1)}\right) \tag{5}$$

It is noticed that, when the characteristic length of sample is in microscale, $e^{mL} \gg 1$. Then, the right side of Eq. 5 can be simplified as

$$\Delta T = \frac{q}{\kappa m^2} \left(1 - \frac{2}{mL}\right) \tag{6}$$

The error made from this simplification is within 0.001%.

Moreover, the resistance change caused by the average temperature change can be described with resistance change rate with temperature $\frac{dR}{dT}$[28]

$$\Delta R = \frac{dR}{dT} \Delta T = \frac{dR}{dT} \frac{q}{\kappa m^2} \left(1 - \frac{2}{mL}\right) \tag{7}$$

The resistance change rate with temperature $\frac{dR}{dT}$ can be calculated using temperature coefficient of resistance $\beta$ shown as Eq. 8[28].

$$\frac{dR}{dT} = \beta R_0 \tag{8}$$

Where $R_0$ is the resistance of the microwire at the environment temperature, and it can be calculated as[28]

$$R_0 = \rho \frac{L}{A} \tag{9}$$

Where $\rho$ is the electrical resistivity.

The heat generated per unit volume $q$ is calculated using Joule's first law

$$q = \frac{I^2 R}{V} = \frac{I^2 R}{AL} \approx \frac{I^2 R_0}{AL} = \frac{I^2 \rho \frac{L}{A}}{AL} \tag{10}$$

Where $V$ is the volume of the wire sample.

Then Eq. 7 can be expressed as:

$$\begin{cases} \Delta R = aL - b \\ a = \frac{\beta I^2 \rho^2}{A^3 \kappa m^2} \\ b = \frac{2\beta I^2 \rho^2}{A^3 \kappa m^3} \end{cases} \tag{11}$$

It shows that there is a linear relationship between the increase of resistance and the length of samples.

Based on Eq. 11, the main equations of MSSHW method are derived out as

$$h = \frac{16\beta I^2 \rho^2}{\pi^3 d^5 \cdot a}$$ (12)

where $d$ is the diameter of the thin wire and $I$ is the given current. That is, convection heat transfer coefficient (h) can be obtained by measuring linear relationship between $\Delta R$ and L without knowing thermal conductivity.

Based on Eq. 11, the thermal conductivity can also be calculated as

$$\kappa = \frac{16\beta I^2 \rho^2 \cdot b^2}{\pi^3 d^6 \cdot a^3}$$ (13)

However, the equation of $\kappa$ includes the intercept b which is a very small value compared with $\Delta R$. The value of $b$ is so small that it is highly uncertain. So that the thermal conductivity $\kappa$ calculated by this equation has a high uncertainty. Therefore, we don't discuss the thermal conductivity here.

To proof the validity of MSSHW method, we measured the natural convection heat transfer coefficient of platinum microwires. Firstly, the diameters of the platinum wire were measured by optical microscope and the picture is shown in Fig. 3b. By averaging the nine measured values at different positions of the microwire, the diameter of the platinum wire is obtained as 41 μm.–Secondly, by measuring their resistances, the lengths of the samples are obtained based on Eq. 9 [28]. Thirdly, the increase of electric resistance ($\Delta R$) of seven samples, whose lengths range from 20 mm to 130 mm, were measured under the current of 60 mA (shown in Fig. 4a). Finally, in accordance with Eq. 11, the parameters of a and b can be obtained by linear fitting. When the current was applied as 60 mA, a and b were obtained as 2.41±0.02 Ω/m and 0.015±0.002 Ω, respectively. Therefore, convection coefficient can be calculated out ~~at a given direct current~~ by Eq. 12 as 246±6 W/m$^2$-K.

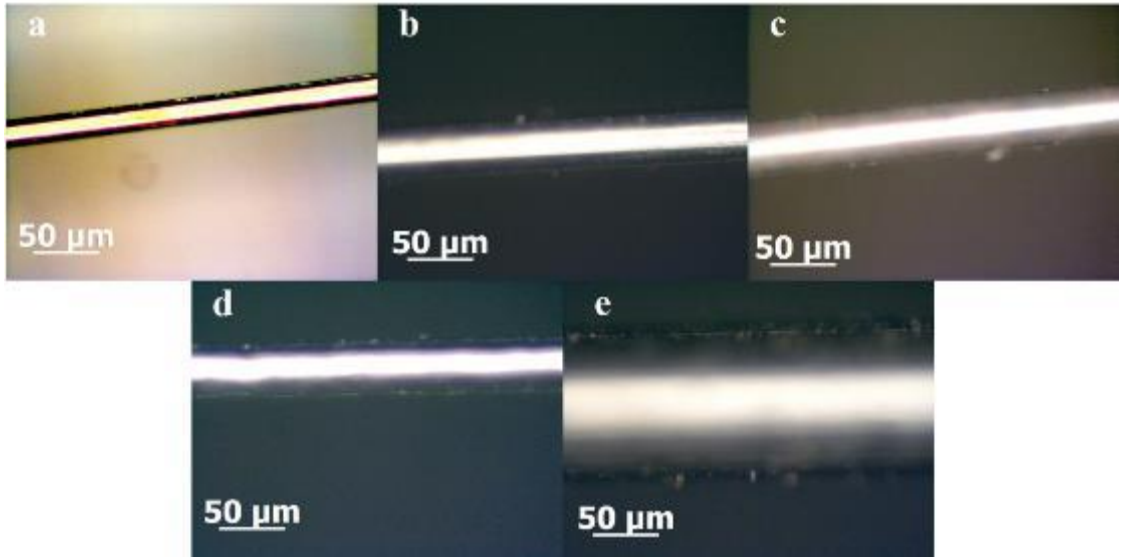

**Fig.3** The microscope photo of microwires (the diameter is measured from at least nine different positions of those microwires): (a) platinum: 20±1 μm in diameter; (b) platinum: 41±3 μm in diameter; (c) Tungsten: 40±1 μm in diameter; (d) stainless steel: 42±2 μm in diameter. (e) stainless steel: 120±6 μm in diameter.

The convection heat transfer coefficient, measured under different direct current, are shown in Fig. 4b. As the results show, the convection coefficient is current independent. Of course, the current can't be too high or too low. The error of measurement will be too high due to the too low increase of average temperature or the radiation heat loss can not be ignored due to the too high increase of average temperature. Our results of convection coefficient agree well with values calculated by the empirical equations of Churchill, Morgan and Fujii. However, our results have a slight difference between Jan's empirical equation[29]. The difference can be attributed to the following factors: thermal conduction to the supports and the temperature measurement locations; distortion of the temperature and velocity fields by bulk fluid movements; the use of undersized containing chambers or the presence of the temperature system and supports; and temperature loading effects[32].

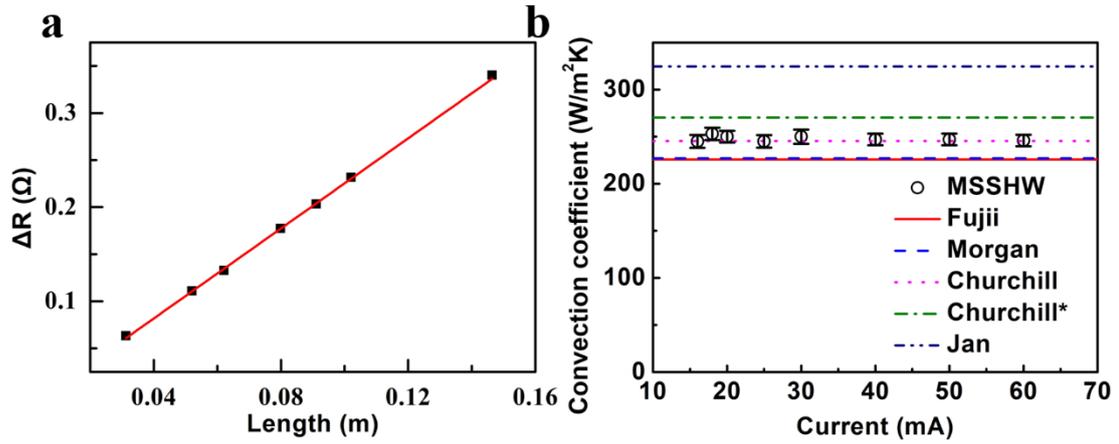

**Fig.4** (a) The increase of electric resistance with length. A 60 mA current is applied in measurement. Red line represents the fitting line. (b) Experimental results of convection heat transfer coefficient with variable current. The black open circles represent values of convection coefficient results from our work (MSSHW method). The red solid line, blue dash line, magenta dot line, green dash-dot line and navy dash-dot-dot line are convection heat transfer coefficient calculated from the empirical equations of Fujii[30], Morgan[31], Churchill[32], Churchill*[32] and Jan[33], respectively.

To study the material effect on the convection heat transfer coefficient, microwires (~40 μm in diameter) of platinum, tungsten, and stainless steel (Fig.3 b, c, d) are measured with MSSHW method. As shown in Fig. 5a, the values of convection heat transfer coefficient are independent on materials. The values of three materials with 40 μm diameter are close to each other, which is also close to the prediction values in Ref.[30-32] by empirical equations. According to the optical microscope images, the slight difference in convection coefficients may be attributed to the surface roughness which has influence on heat convection[34].

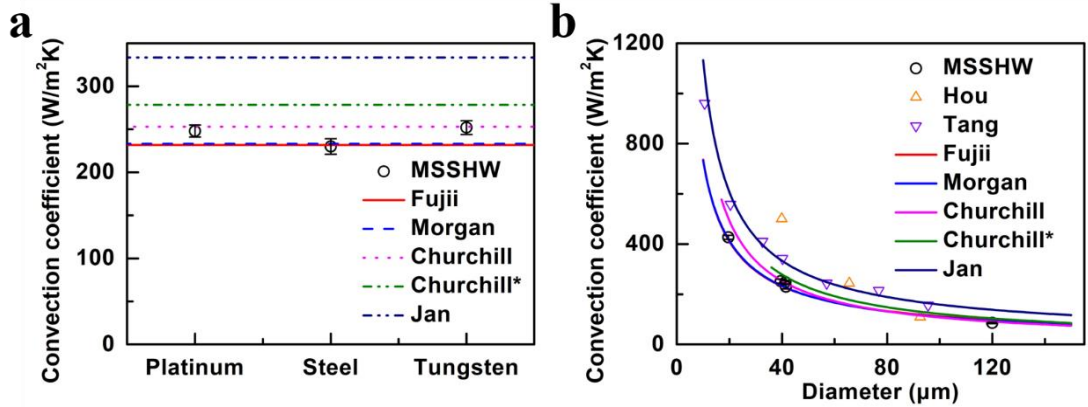

**Fig.5** (a) The natural convection heat transfer coefficients of the three microwires with different materials measured by MSSHW method. The diameter of platinum, stainless steel and tungsten microwires is 41 μm, 42 μm and 40 μm, respectively. (b) Experimental convection heat transfer coefficients under different diameters measured by MSSHW method. (black open circles were measured results by MSSHW method, the orange up triangles were measured by Hou[25], and the purple down triangles were measured by Tang[24].)

Moreover, microwires with different diameters were measured to show the size effect of convection heat transfer coefficient (Fig. 5b). As we can see, the diameter largely affects convection heat transfer coefficient. With diameter decreasing from 120 μm to 20 μm, the convection heat transfer coefficient increases from 86 W/(m² · K) to 427 W/(m² · K). The measured convection heat transfer coefficients decrease sharply when the diameter increases, and converge to a constant for bulk. This phenomenon might attribute to two physical mechanisms[35]. Firstly, when the diameter of the wire is comparable to the mean free path of the air molecular, the continuum theory breaks. Thus, the theory for the macroscopic phenomena is no longer applicable. Secondly, the surface volume ratio increases with the decreases of wire diameter, making the surface-related influences more significant [35]. In comparison, both experimental results[24,25] and empirical equations results[30-33] are also shown in Fig. 5b.

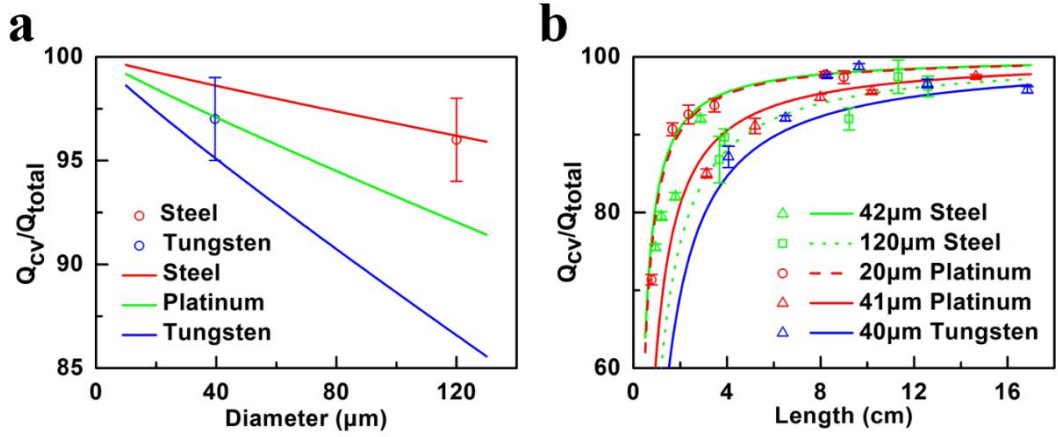

Fig.6 (a) The percentage of convection heat transfer under different diameters. The length of those microwires is ~126mm (dots are measured results by MSSHW method and calculated by Eq. 14; lines are calculated by Eq.15 with parameters from Ref.[31]). (b) The percentage of convection heat transfer under different lengths.

To get a better understanding of heat transfer in microscale, the percentage of heat transferred by convection ($Q_{cv}/Q_{total}$) has been investigated. Based on Eq.2 and Eq.10, $Q_{cv}/Q_{total}$ can be obtained by measurements as

$$Q_{cv}/Q_{total} = \frac{hPL\Delta T}{I^2R} \tag{14}$$

Besides, from Eq. 5, the value of ratio can also be estimated as

$$Q_{cv}/Q_{total} = \frac{hPL\Delta T}{qAL} = 1 - \frac{2}{mL}\left(\frac{e^{mL}-1}{e^{mL}+1}\right) \tag{15}$$

Firstly, we studied the dependence of $Q_{cv}/Q_{total}$ on the diameter of microwires. The data from measurements (Eq. 14) are matched well with the prediction curves by theory (Eq.15). As shown in Fig. 6a, with the diameter decreasing, $Q_{cv}/Q_{total}$ increases. For example, for a steel microwire with a diameter of 120 μm and a length of 126mm, the percentage of heat flux carried by convection is more than 95%. So, convection plays a predominant role in heat transfer of microwires.

Moreover, the ratio depends on the materials, as shown in Fig. 6a. The thermal convections of microwires are not sensitive on materials. However, the thermal conduction depends on materials. The thermal conductivity of steel, platinum and tungsten microwires measured by traditional steady state hot wire method are 18±3, 79±3 and 224±9 W/mK, respectively. Because the steel microwire has a lower value of thermal conductivity than platinum and tungsten. it has a higher value of $Q_{cv}/Q_{total}$.

Secondly, we studied the dependence of $Q_{cv}/Q_{total}$ on the length of microwires. Fig. 6b shows that, with the increasing of length, the ratio of $Q_{cv}/Q_{total}$ increases. It means that the longer a wire is, the more important the convective heat transfer is. Because the area of convection increases with the length.

**CONCLUSION**

In conclusion, a modified steady-state hot wire (MSSHW) method is proposed to characterize the conductive and convective heat transfer of microwires simultaneously. The method is verified by measuring the convection heat transfer coefficient of microwires. The convective heat transfer between micro metal wires and their surrounding air environment is found to be irrelevant to material composition, but is strongly connected with the diameter of the wires. When the diameter of microwires decreases from 120 μm to 20 μm, the natural convection coefficient increases from 86 W/(m$^2 \cdot$K) to 427 W/(m$^2 \cdot$K). The convection coefficient is in reasonable range compared with those from the references. This MSSHW method provides a convenient way for measuring convective heat transfer of microwires without knowing thermal conduction, which is beneficial to studying the comprehensive heat transfer at microscale

**ACKNOWLEDGEMENTS**

This work is financially supported by the National Natural Science Foundation of China (No. 51576076, No. 51711540031 and No. 51622303), the Natural Science Foundation of Hubei Province (No. 2017CFA046), and the Fundamental Research Funds for the Central Universities (No. 2016YXZD006). The authors thank the National Supercomputing Center in Tianjin (NSCC-TJ) and China Scientific Computing Grid (ScGrid) for providing assistance in computations. We are grateful to Xiaoxiang Yu for useful discussions. We also acknowledge the precious suggestions about this paper from Wei Liu, Zhichun Liu, Haisheng Fang and Run Hu.